\documentclass[a4paper,11pt]{article}
\pdfoutput=1
\input{paper.sty}
\usepackage{jcappub}

\definecolor{dred}{RGB}{163,35,47}
\definecolor{dblue}{RGB}{45,90,195}

\newcommand{\pd}{\partial}
\newcommand{\dd}{\mathrm{d}}
\newcommand{\avg}[1]{\left\langle#1\right\rangle}

\newcommand{\al}{\alpha}
\newcommand{\be}{\beta}

\newcommand{\de}{\delta}
\newcommand{\la}{\lambda}

\newcommand{\ep}{\epsilon}

\newcommand{\ta}{\theta}
\newcommand{\om}{\omega}
\newcommand{\Om}{\Omega}
\newcommand{\io}{\iota}

\newcommand{\vpi}{\varpi}

\newcommand{\vth}{\vartheta}
\newcommand{\vph}{\varphi}

\newcommand{\der}{{\de r}}
\newcommand{\denu}{{\de\omega}}
\newcommand{\nnut}{{n \omega t}}
 
\newcommand{\Dc}{\Delta c}
\newcommand{\dg}{\(^\dag\)}
\newcommand{\ddg}{\(^\ddag\)}
\newcommand{\cO}[1]{\mathds{O}\left(#1\right)}

\newcommand{\rhoDM}{\rho_{\text{DM}}}

\newcommand{\deq}{\coloneqq}

\newcommand{\Eq}[1]{Eq.~(\ref{#1})}
\newcommand{\Eqs}[2]{Eqs.~(\ref{#1}) and~(\ref{#2})}

\newcommand{\nn}{\nonumber} 

\newcommand{\te}{\text{\itshape{ə}}}

\usepackage{newfloat}
\DeclareFloatingEnvironment[
    fileext=lop,
    listname={List of Panels},
    name=Panel,
    placement=tbhp,
    within=none,
]{panel}

\title{\textcolor{dblue}{Vector Fuzzy Dark Matter, Fifth Forces, and Binary Pulsars}}

\author[a]{Diana L\'opez Nacir,}
\author[b]{and Federico R.~Urban}

\affiliation[a]{Departamento de F\'isica and IFIBA, FCEyN UBA, Facultad de Ciencias Exactas y Naturales\\Ciudad Universitaria, Pabellon I, 1428 Buenos Aires, Argentina}
\affiliation[b]{CEICO, Institute of Physics of the Czech Academy of Sciences\\Na Slovance 2, 182 21 Praha 8, Czech Republic}

\abstract{We study the secular effects that an oscillating background ultralight (fuzzy) cosmological vector field has on the dynamics of binary systems; such effects appear when the field and the binary are in resonance.  We first consider the gravitational interaction between the field and the systems, and quantify the main differences with an oscillating background scalar field.  If the energy density of such a field is sufficiently large, as required if it is supposed to be all of the dark matter, we show that the secular effects could yield potentially observable signatures in high precision time of arrival measurements of binary pulsars.  We then analyse the secular effects that arise when the field is directly coupled to the bodies in the binary.  We show that this study is particularly relevant for models where fuzzy dark matter mediates a baryonic force \( B \) (or \( B-L \), with \( L \) the lepton number), due to the stellar amount of nucleons present in the stars.  The constraints we obtain from current data are already competitive with (or even more constraining than) laboratory tests of the equivalence principle.}

\begin{document}
\maketitle
\flushbottom

%===============================================================================
% BODY
%===============================================================================

%-------------------------------------------------------------------------------
\section{Motivation}
\label{sec:motivation}
%-------------------------------------------------------------------------------

Cosmological Dark Matter (DM) is one of the standard ingredients in models of the Universe: accounting for roughly 27\% of the total energy density, it is more than five times in density than visible baryonic matter~\cite{Aghanim:2018eyx}.  All observational evidence for DM is due to its direct or indirect gravitational effects~\cite{Bertone:2004pz}.  While alternatives exist~\cite{Deffayet:2011sk,Deffayet:2014lba,Berezhiani:2015bqa,Berezhiani:2017tth}, most efforts so far have focussed on particle DM, in which one (or more) new fields are introduced as extensions of the Standard Model, to account for the missing matter density.  However, so far all DM production or detection experiments have yielded null results, and the nature of this field remains as elusive as ever~\cite{Akerib:2016vxi,Cui:2017nnn,Aprile:2018dbl}.  With the traditional candidates nearly ruled out by non-detection, before turning to the implausible, it is important to explore all plausible possibilities.

One such option is that DM is in the form of an extremely light (fuzzy) particle, which during the late-time evolution of the Universe behaves as an oscillating classical field; such oscillations make the energy momentum tensor of the field look like dust, hence DM, in a cosmological setting, see~\cite{Preskill:1982cy,Turner:1983he,Hu:2000ke,Nelson:2011sf,Marzola:2017lbt,Marsh:2015xka}.  In this paper we explore the possibility that fuzzy DM is a vector, and we will show that, in the range of masses \( 10^{-23}\text{eV}\lesssim m \lesssim10^{-18}\text{eV} \)\footnote{Lower masses are excluded due to their effect on structure formation, and in fact the lower end of the range we consider could be in tension with data, see~\cite{Irsic:2017yje,Armengaud:2017nkf,Zhang:2017chj}.  Complementary tests (which apply even if the field is not the DM) can be obtained from rotating Black Holes systems (with current data already disfavoring fields with \(m\sim 5\times10^{-14}-2\times 10^{-11}\)eV~\cite{Baryakhtar:2017ngi} or binary Black Holes~\cite{Baumann:2018vus}.}, precision timing measurements of binary pulsars offer a unique possibility to test the properties of such DM candidate, as was observed in~\cite{Blas:2016ddr} for the case of scalar fuzzy DM.  Indeed, \emph{when the DM oscillations are in resonance with the binary system}, the former causes a secular variation of the orbital parameters of the latter.

Here we obtain two separate results.  First, based on gravitational interactions alone, we derive the qualitative differences in the peculiar distortions imprinted on binary systems by vector DM compared to scalar DM.  If measured, these differences would allow one to infer the spin of the hypothetical fuzzy DM field.  Second, already with current data, if vectorial fuzzy DM is a carrier of a baryonic force, we can place bounds on its strength which are competitive with (or even more restrictive than) those obtained by equivalence principle laboratory tests.

The paper is organised as follows.  In Section~\ref{sec:action} we describe the model and its cosmology.  We then proceed to the study of binary systems in the case in which DM and the binary interact only gravitationally (Section~\ref{sec:gravity}), and to the case when a fifth force is present due to direct interactions (Section~\ref{sec:fifth}).  We discuss our results and perspectives in Section~\ref{sec:end}.  The Appendix contains a brief review of the osculating orbits method to discuss secular variations of keplerian orbits, the full system of secular variations of orbital parameters, and the list of the binary systems we analysed (with their properties).

%-------------------------------------------------------------------------------
\section{Vector fuzzy DM}
\label{sec:action}
%-------------------------------------------------------------------------------

The main ingredient of the model is a Proca vector described by the action:
\begin{align}\label{eq:action}
  S \deq - \int\!\dd^4x \sqrt{-g} \left[ \frac14 F^{\mu\nu} F_{\mu\nu} - \frac12 m^2 A^\mu A_\mu \right] \,,
\end{align}
where \( F_{\mu\nu} = 2 \pd_{[\mu} A_{\nu]} \) is the \( A_\mu \) vector field strength and \( m \) its mass; we use the mostly negative signature for the metric, and we define symmetrisation and antisymmetrisation as \( 2T_{(\mu\nu)} \deq T_{\mu\nu} + T_{\nu\mu} \) and \( 2T_{[\mu\nu]} \deq T_{\mu\nu} - T_{\nu\mu} \), respectively.  The energy momentum tensor (EMT) is defined by
\begin{equation}
  \de S \deq \int\!\dd^4x \sqrt{-g} \tau_{\mu\nu} \de g^{\mu\nu} \,,
\end{equation}
which explicitly reads
\begin{align}\label{eq:emt1}
  \tau_{\mu\nu}	=	&\, - F_\mu^{~\la} F_{\la\nu} + \frac14 g_{\mu\nu} F^{\al\be} F_{\al\be} + m^2 (A_\mu A_\nu - \frac12 g_{\mu\nu} A^\la A_\la) \,.
\end{align}
To study the cosmology of this theory we choose the Lorenz gauge \( \pd_i A^i = 0 = A_0 \), and focus on the Friedmann-Lemaître-Robertson-Walker metric: \( \dd s^2 = \dd t^2 - a(t)^2 |\dd\vec{x}|^2 \) with \( a(t) \) the scale factor of the Universe.  The equations of motion for the homogeneous (zero) mode \( \vec{A} = \vec{A}(t) \) are
\begin{align}
  & \ddot{A}_i + H\dot{A}_i + m^2 A_i = 0 \,, \label{eq:eom1}\\
  & \ddot{A}^i + 5H\dot{A}^i + 2(\dot{H}+3H^2) A^i + m^2 A^i = 0 \,, \label{eq:eom2}
\end{align}
where \( H \deq \dd\log{a}/\dd t \) is the Hubble parameter, and an overdot stands for cosmic time partial derivative.

The solutions to a differential equation of the type
\[
  \ddot{f} + p H\dot{f} + m^2 f = 0
\]
with \( H = H_0 a^{-q} = 1/qt \) and \(p\) and \(q\) numerical constants, is
\[
  f(t) = a^{(q-p)/2} \left[ C_J J_{\frac{p}{2q}-\frac{1}{2}}(mt) + C_Y Y_{\frac{p}{2q}-\frac{1}{2}}(mt) \right] \,,
\]
with \( J \) and \( Y \) Bessel functions and \( C_J \) and \( C_Y \) arbitrary constants.  In the case of fuzzy vector DM we obtain, upon expanding for a rapidly oscillating vector \( mt \gg 1 \)
\begin{align}\label{eq:vector}
  A_i(t) =	&\, a^{-1/2} \left[ \check{C}_J \sin(mt) + \check{C}_Y \cos(mt) \right] \nn\\
  \deq		&\, \hat{A}_i a^{-1/2} \cos(mt+\Upsilon) \,,
\end{align}
in matter domination (here \( \check{C}_J \) , \( \check{C}_Y \) and \( \Upsilon \) are arbitrary constants)\footnote{Notice that we have \( A^i = - \hat{A}^i a^{-5/2} \cos(mt+\Upsilon) \), that is, the \( \hat{A}^i \deq \hat{A}_i \) are not components of a vector.}.

In the cartesian orbital \( (x,y,z) \) frame associated to a given binary system (see Fig.~\ref{fig:orbits}), the vector field can be written in terms of spherical coordinates \( (\vth,\,\vph) \)  
\begin{align}\label{eq:a_def}
  \hat{A}_i \deq \hat{A} (s_\vth c_\vph, s_\vth s_\vph, c_\vth) \,,
\end{align}
where we employ the shortcut notation 
\begin{subequations}\label{shortcut}
  \begin{align}
  s_x&\deq\sin x \,,\\
  c_x&\deq\cos x \,; 
    \end{align}
\end{subequations} the EMT then becomes
\begin{subequations}
  \begin{align}
    \tau^0_0 =	&\, \frac12 \hat{A}^2 m^2 a^{-3} \left[ 1 + \cO{\frac{H}{m}} \right] \,,\label{eq:t00ex}\\
    \tau^0_j =	&\, 0 \,,\label{eq:t0iex}\\
    \tau^i_{~j} =	&\, \frac12 \hat{A}^2 m^2 a^{-3} \left[ c_{2( mt+ \Upsilon)} + \cO{\frac{H}{m}} \right] \hat{X}^i_{~j} \,,\label{eq:tijez}
  \end{align}
\end{subequations}
where \( \hat{X}^i_{~j} \deq \de^i_{~j} - 2X^i_{~j} \) and \( X^i_{~j} \deq \hat{A}^i \hat{A}_j^T / \hat{A}^2 \deq a^i a_j^T \).  In the late-time Universe, when \( m/H\gg1 \), the field oscillates rapidly, and upon averaging over a Hubble time the EMT reduces to \( \avg{\tau^\mu_\nu} \deq \mathrm{diag}(\rhoDM,\vec{0}) \), with \( \rhoDM \deq \hat{A}^2 m^2 / 2a^3 \) the energy density of the vector zero mode\footnote{From this point onward we will not need the scale factor \( a \) as pulsar timing measurements are obviously taken at \( a\approx1 \); we will use this symbol for the orbital semimajor axis from now on, see Fig.~\ref{fig:orbits}.}.  The pressure term is therefore suppressed, and can be seen as a small perturbation on the background governed by \( \rhoDM \):
\begin{align}\label{eq:emt_fl}
  \tau^{(1)i}_{~~~j} = \rhoDM c_{2(mt+\Upsilon)} \hat{X}^i_{~j} \deq \de^i_{~j} p + \Pi^i_{~j} \,,
\end{align}
where \( p = \rhoDM c_{2(mt+\Upsilon)} /3 \) is the isotropic pressure and \( \Pi^i_{~j} = -2\rhoDM c_{2(mt+\Upsilon)} (X^i_{~j}-\de^i_{~j}/3) \) the traceless anisotropic stress \( \Pi^i_{~i} = 0 \); both of them are (first order) \emph{perturbations} compared to \( \rhoDM \).

%-------------------------------------------------------------------------------
\section{Gravitational interaction}
\label{sec:gravity}
%-------------------------------------------------------------------------------

%-------------------------------------------------------------------------------
\subsection{Gravitational perturbations}\label{ssec:pert}
%-------------------------------------------------------------------------------

The effect of the oscillating background perturbation~\Eq{eq:emt_fl} on the binary system is encoded in the force per unit mass
\begin{align}\label{eq:pertb}
  F^i \deq \ddot{\der}^i = R^{(1)i}_{~~~0j0} r^j \,,
\end{align}
in Fermi normal coordinates associated with the centre of mass of the system \( r^i \).  To derive the perturbed Riemann tensor we expand the metric around a Minkowski background as \( g_{\mu\nu} \deq \eta_{\mu\nu} + g^{(1)}_{\mu\nu} \) and
\begin{subequations}
  \begin{align}
    g^{(1)}_{00} \deq	&\, 2\Phi \,, \label{eq:g00}\\
    g^{(1)}_{0i} \deq	&\, - \pd_i B + S_i \,, \label{eq:g0i}\\
    g^{(1)}_{ij} \deq	&\, 2\de_{ij}\Psi - 2\pd_i\pd_jE + \pd_{(i}F_{j)} +  h_{ij} \,, \label{eq:gij}
  \end{align}
\end{subequations}
with, as usual, \( \pd^i S_i = 0 = \pd^i F_i \) and \( \pd_i h^i_{~j} = 0 = h^i_{~i} \).  In the Poisson gauge\footnote{Clearly our results are gauge-independent.} \( B = E = 0 = F_i \) the perturbed Einstein equations
\begin{align}\label{eq:einstein}
  G^{(1)\,i}_{~~~~j}														= 		&\, 8\pi G T^{(1)\,i}_{~~~~j} \nn\\
																		\Downarrow	&\, \nn\\
  - 2\de^i_{~j}\ddot{\Psi} - \pd^{(i}\dot{S}_{j)} - \frac12 \ddot{h}^i_{~j}	=		&\, 8\pi G \left[ \de^i_{~j} p + \Pi^i_{~j} \right] \,,
\end{align}
where \( G \) is the Newton's constant, give \( \ddot{\Psi} = - 4\pi G p \) and \( \pd^{(i}\dot{S}_{j)} + \frac12 \ddot{h}^i_{~j} = - 8\pi G \Pi^i_{~j} \) (we have neglected scalar and tensor gradients).  Therefore, the Riemann tensor turns out to be
\begin{align}\label{eq:riemann}
  R^{(1)i}_{~~~~0j0} = &\, - \de^i_{~j}\ddot{\Psi} + \pd^{(i}\dot{S}_{j)} + \frac12 \ddot{h}^i_{~j} = 4\pi G \rhoDM \left[ 4X^i_{~j} - \de^i_{~j} \right]  c_{2(mt+\Upsilon)}  \,.
\end{align}

%-------------------------------------------------------------------------------
\subsection{Results}\label{ssec:grav_res}
%-------------------------------------------------------------------------------

The binary Keplerian orbits are described by the Lagrange planetary equations (see the Appendix), which are given directly in terms of the force per unit mass \Eq{eq:pertb}.  We focus here on the variation of the orbital period, \( \dot{P}_b\), as it is the parameter that gives the stronger constraints, and we refer to the Appendix for the full system of the six independent parameters.  The orbital period can be directly obtained from the (variation of the) semi-major axis \( a \) thanks to
\[
  P_b\deq\frac{2\pi}{\omega_0}=2\pi\sqrt{\frac{a^3}{GM_T}} \,,
\]
where \( \omega_0 \) is the orbital frequency, and \( M_T \deq M_1 + M_2 \) is the total mass of the system.

\begin{figure}[tbhp]
  \center{\includegraphics[width=8cm]{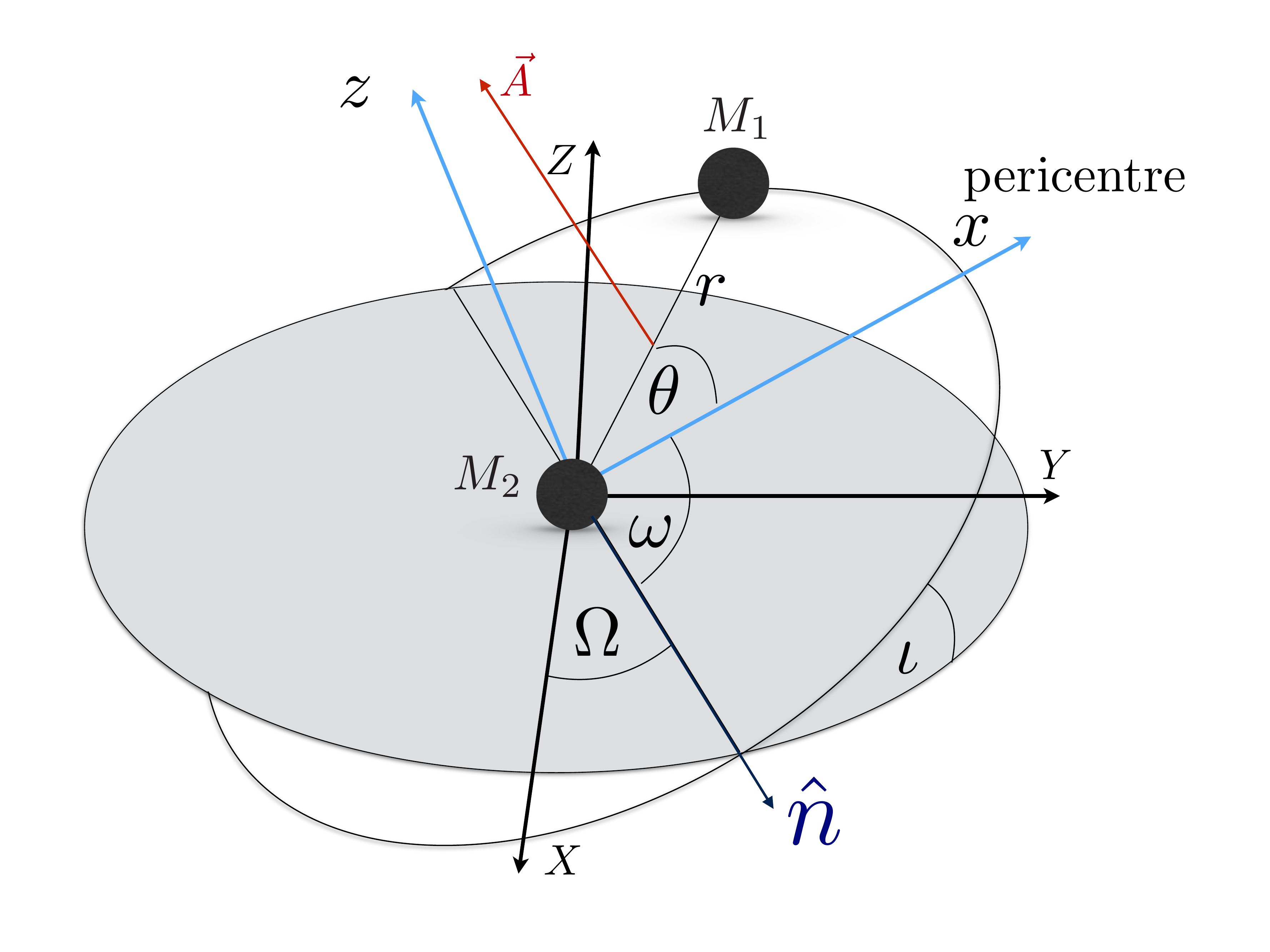}}
  \caption{Description of Keplerian orbits in terms of the orbital elements viewed in the fundamental \( (X,Y,Z) \) reference frame.  The cartesian orbital \( (x,y,z) \) frame and the polar one \( (r,\ta,z) \) are also shown (centered at \( M_2 \) for convenience).  Here \( \vec{A} \) stands for the vector background field describing DM, \( \hat{n} \) is the unit vector pointing towards the ascending node.}\label{fig:orbits}
\end{figure}

The variation of \( a \) is given by
\begin{align}
  \frac{\dot{a}}{a}	= &\, \frac{2}{\omega_0}\left\{\frac{e\sin\ta}{a\te} F_r + \frac{\te}{r} F_\ta \right\} \,, \label{eq:a}
\end{align}
where we have decomposed the vector \( \vec{F} \) in the reference frame of the binary system with polar coordinates \( (r,\ta,z) \) as \( \vec{F}=F_r\hat{r}+F_{\ta}\hat{\ta}+F_z\hat{z} \), see Fig.~\ref{fig:orbits}, and we defined \( \te\deq\sqrt{1-e^2} \).  Explicitly, in terms of the perturbation in \Eq{eq:riemann} we obtain
\begin{subequations}
  \begin{align}
    F_r		= &\, 4\pi G \rhoDM c_{2(mt+\Upsilon)} \, r \left[c_{2(\ta-\vph)}-2c_{2 \vth} c_{\ta-\vph}^2\right] \,, \label{eq:fr}\\
    F_\ta	= &\, -8\pi G \rhoDM c_{2(mt+\Upsilon)}\, r \, s^2_\vth s_{2(\ta-\vph)} \,, \label{eq:fta}\\
    F_z		= &\, 8\pi G \rhoDM c_{2(mt+\Upsilon)} \, r \, s_{2\vth} c_{(\ta-\vph)} \,.\label{eq:fz}
  \end{align}
\end{subequations}

As we show next, we obtain secular variations of the orbital parameters when the binary system is in resonance with the oscillating background.  The first step is to express the orbit in terms of Bessel series in \( \sin[n\omega_0(t-t_0)] \) and \( \cos[n\omega_0(t-t_0)] \), with \(t_0\) the time of periastron.  If we parametrise the (small) gap between the two frequencies as \( \denu \deq 2m - N\omega_0 \), where \( N \) is the resonance harmonic number, then, upon averaging over a long time \( \Delta t \) for which \( P_b \ll \Delta t \ll 2\pi/\denu \):
\[
  \avg{f(t)} \deq \frac{1}{\Delta t} \int_t^{t+\Delta t}\dd t \, f(t) \,,
\]
we have
\begin{align}
  \avg{s_{n\omega_0(t-t_0)}c_{2(mt +\Upsilon)}} &\approx -\frac{1}{2}\de(n-N)s_{\gamma_g(t) }\,, \nn\\
  \avg{c_{n\omega_0(t-t_0)}c_{2(mt	 +\Upsilon)}}  & \approx  \frac{1}{2}\de(n-N)c_{\gamma_g(t)}\,, \nn
\end{align} with
\begin{align}
  \gamma_g(t) \deq \denu(t-t_0)+2mt_0+2\Upsilon \,.
\end{align}

Therefore, the \(\de(n-N)\) selects the \(n=N\) term in the Bessel series, and this is the secular contribution to the variation of the orbital parameters.  Keeping only this dominant secular term, we then obtain:
\begin{align}\label{eq:dTdt_avg}
  \avg{\dot{P}_b} = \frac32 G \rhoDM P_b^2	&\, \left\{ \left[Q_{xx}(Ne)+Q_{yy}(Ne)\right] s_{\gamma_g(t)} + Q_{xy}(Ne) c_{\gamma_g(t)} \right\} \,,
\end{align}
where
\begin{align}\label{eq:QforT}
  \begin{cases}
    Q_{xx}(Ne) = \left(4 s_\vth^2 c_\vph^2 - 1 \right) N q_{xx}(Ne) \\
    Q_{yy}(Ne) = \left(4 s_\vth^2 s_\vph^2 - 1 \right) N q_{yy}(Ne) \\
    Q_{xy}(Ne) = 4 s_\vth^2 s_{2\vph} N q_{xy}(Ne)
  \end{cases} \,,
\end{align}
and the \( q(Ne) \) are defined in \Eq{eq:qdef} in the Appendix.

%-------------------------------------------------------------------------------
\subsection{Phenomenology}\label{ssec:grav_pheno}
%-------------------------------------------------------------------------------

As expected from the  symmetry of the system, when the vector is directed along \( \hat{z} \) we recover exactly the scalar result; however, the vector case has a much richer phenomenology.  For instance, the difference with the scalar is most evident if the orbits are exactly circular: in this case the scalar secular drift disappears, whereas the vector gives a non-zero effect, since \( q_{xx}(0) = - q_{yy}(0) = q_{xy}(0) = 1/2 \) (when \( N=2 \) only).  We can write \Eq{eq:dTdt_avg} in this case as
\begin{align}\label{eq:dTdt_zero}
  \avg{\dot{P}_b}	& \to 6G\rhoDM P_b^2 s_\vth^2 s_{\gamma_g(t)+2\vph} \simeq 1.6\times 10^{-21} \left(\frac{P_b}{ \text{d}}\right)^{2} \text{s} \, \text{s}^{-1} \,,
\end{align}
where \({\text{d}}\) stands for days, we have assumed \( s_\vth^2 s_{ \gamma_g(t)+2\vph} = 1 \) and used  \( \rhoDM = 0.3 \mbox{GeV}/\mbox{cm}^3 \) as a typical value for the local DM density. 

For known binary systems, as can be seen in Table~\ref{tab:frac}, the typical errors on \( \dot{P}_b\) are at best of \(\cO{10^{-15}}\)\footnote{For the double pulsar PSR J0737-3039, although work on this system is still ongoing, an accuracy of \(\cO{10^{-16}}\) is expected already from current data~\cite{Kehl:2016mgp}.}.  Therefore, the effect described above is too small to affect current measurements of \(\dot{P}_b\).  In the future, one can expect the error on \(\dot{P}_b\) to improve roughly as \(\sim \left({T_0}/{T}\right)^{5/2} \left({\delta t}/{\delta t_0}\right)\), where \(T\) (\(T_0\)) is the future (current) observational time and \(\delta t\) (\(\delta t_0\)) the future (current) time of arrival (TOA) precision~\cite{Teukolsky1976,Damour:1991rd}.  For instance, if recently discovered, a factor \(10^{5/2}\) of improvement can be expected after observing the system for 10~years.  The TOA precision can be improved by an order of magnitude, \({\delta t }/{\delta t_0}\sim10\), with the next generation of radio telescopes, such as the Square Kilometre Array (SKA)\footnote{https://www.skatelescope.org}, see for instance~\cite{Liu:2011cka}.

Improvements by more than an order of magnitude, while possible, are in general very difficult to achieve.  Notice that while the effect increases for systems with long orbital periods, scaling as \(P_b^2\), the precision worsens according to \(P_b^{4/3}\).  Therefore, measuring this effect will be challenging even for future experiments.  Lastly, better chances can be had in denser environments such as closer to the Galactic centre, where \(\rhoDM\) could be a factor~10 or more higher, so that the effect would be boosted and measureable.

Before closing this section, note that, in order for the secular drift to appear, the resonance should be sustained across many binary periods.  Since all expected (theoretical) General Relativity effects, as well as all empirically measured \(\dot{P}_b\) are minuscule \(\dot{P}_b\ll1\), this assumption is accurately satisfied.  Lastly, the vector field should also retain its direction and phase for many binary periods.  Given that the expected coherence time is or order \(v^2/m\) and that the typical velocity of a virialised halo is \(v\sim10^{-3}\), we expect this to be true for roughly \(10^6\) periods --- for a period of \(P_b\approx1\)d this means almost 3000 years.

%-------------------------------------------------------------------------------
\section{Direct coupling and fifth forces}
\label{sec:fifth}
%-------------------------------------------------------------------------------

%-------------------------------------------------------------------------------
\subsection{Theory}\label{ssec:fifth_thy}
%-------------------------------------------------------------------------------

In addition to the unavoidable gravitational coupling, the vector fuzzy DM could interact non-gravitationally to the binary.  There are two possibilities.  In the first case the fuzzy DM field mixes with the standard \( U(1) \) photon~\cite{Nelson:2011sf,Arias:2012az,An:2013yfc,An:2014twa,Dubovsky:2015cca}; this essentially amounts to assigning a (very) small electric charge to the DM, and, as we will see, since a neutron star is practically electrically neutral, this effect is unobservable.

The second option is that DM carries a tiny charge associated with the nucleons of the star, for example baryon number \( B \), or \( (B-L) \) (\( L \) is the lepton number)~\cite{Heeck:2014zfa,Graham:2015ifn,Knapen:2017xzo}.

In all cases we can capture the dynamics by introducing an interaction term
\begin{align}\label{eq:lag-dir}
  L_q \deq q_1 \vec{v}_1 \cdot \vec{A} + q_2 \vec{v}_2 \cdot \vec{A} \,,
\end{align}
where \( q_k \) with \( k \in \{1,2\} \) are the effective charges of the two binary bodies, and \( \vec{v}_k \) are their velocities in the usual \( (x,y,z) \) cartesian orbital reference system.  From the Euler-Lagrange equations we derive the force per unit mass \( F^i_q \):
\begin{align}\label{eq:force-dir}
  F^i_q = - \frac{q} {M_\odot}\dot{A}^i = -\frac{q}{M_\odot} \sqrt{2\rhoDM} s_{mt+\Upsilon}\,a^i \,,
\end{align}
where we have defined \( q/M_\odot \deq (q_1 M_2 - q_2 M_1) / M_1 M_2 \) with \( M_\odot \) the mass of the Sun.

The calculation proceeds in much the same way as for the previous section, except that the perturbation is now given by \Eq{eq:force-dir}.  The result for the change in the semi-major axis is
\begin{align}\label{eq:a-dir}
   \avg{\frac{\dot{a}}{a}}	= &\, \frac{q\sqrt{2\rhoDM}N}{a\omega_0 M_\odot} s_\vth \left\{ q_y s_\vph s_{\gamma_l(t)} - q_x c_\vph c_{\gamma_l(t)} \right\} \,,
\end{align}
where the frequency gap is defined as \( \denu' \deq m-N\omega_0 \) (notice the factor of 2 difference from \( \denu \)), and the time averaging picks up the \( n=N \) terms only:
\begin{align}
  \avg{s_{ n \omega_0 (t-t_0)}s_{m t+\Upsilon}} \approx &\,\frac{1}{2} \de(n-N)c_{\gamma_l(t)}  \,, \nn\\
  \avg{c_{ n\omega_0 (t-t_0)}s_{ m t+\Upsilon}} \approx &\,\frac{1}{2} \de(n-N)s_{ \gamma_l(t)}  \,, \nn
\end{align} with
\begin{align}
  \gamma_l(t) \deq \denu'(t-t_0)+mt_0+\Upsilon \,.
\end{align}

We focus now on the \( B \) and \( (B-L) \) fifth forces and comment on the (unobservable) dark photon case below.  We parametrise the \( B \) or \( (B-L) \) number of the \(k\)-th binary member as \( N_k \deq c_k M_k / m_\text{n} \) with \( m_\text{n} \) the mass of the neutron\footnote{In order for this expression to be valid we need to ensure that the fifth force has a long enough range to ``see'' the entire system, which means that \( m\ll10^{-16}\)eV; this is easily satisfied for all the systems we consider.}; here \( c_k \) is a phenomenological parameter that depends on several factors, most importantly the actual baryonic to gravitational mass ratio, and the proton content of the star~\cite{Prakash:1996xs}; other factors are the compactness of the star and its equation of state, and its gravitational mass itself.  Since here we are interested in an order of magnitude estimate, we employ a typical value of \( \Dc \deq c_1-c_2 \sim 0.1 \), in what follows.  The overall effective coupling can be written as \( q_k \deq g N_k \), where \( g \) is the fifth force strength\footnote{If the DM is not the mediator of fifth baryonic force itself, but is instead coupled to it indirectly through, e.g., a \( Z' \) portal, we can write \( q_k \deq g q_{AZ'} N_k \), with \(q_{AZ'}\) the coupling to the portal.}.

%-------------------------------------------------------------------------------
\subsection{Results and discussion}\label{ssec:fifth_end}
%-------------------------------------------------------------------------------

We start by considering the secular variation of the orbital period from \Eq{eq:a-dir} for nearly circular orbits.  Using the expressions for \(q_x\) and \(q_y\) given in the Appendix as \Eqs{qJexpa1}{qJexpa2}, it is immediate to see that in the limit \(e\to0\), only the first resonance \(N=1\) survives, \(q_x\sim q_y\to1\) and \Eq{eq:a-dir} reduces to
\begin{align}\label{eq:dir_dotp}
  \dot{P}_b	& \to- \frac{3g\sqrt{2\rhoDM}}{2m_\text{n} \left(2\pi GM_T\right)^{1/3}} \, \Dc \, P_b^{4/3} \, s_\vth c_{\gamma_l(t)+\vph} \nn\\
			& \simeq 6.3\times 10^{11} g \Dc \left(\frac{M_\odot}{M_T}\right)^{1/3} \left(\frac{P_b}{\text{d}}\right)^{4/3} \text{s} \, \text{s}^{-1} \,,
\end{align}
where in the second line we have assumed \( s_\vth c_{\gamma_l(t)+\vph} =- 1 \) and \( \rhoDM = 0.3 \mbox{GeV}/\mbox{cm}^3 \).  Higher harmonics (\(N\geq2\)) are therefore only relevant for eccentric systems \(e\geq0.1\).

\begin{figure*}[tbhp]
  \center{\includegraphics[width=0.7\textwidth]{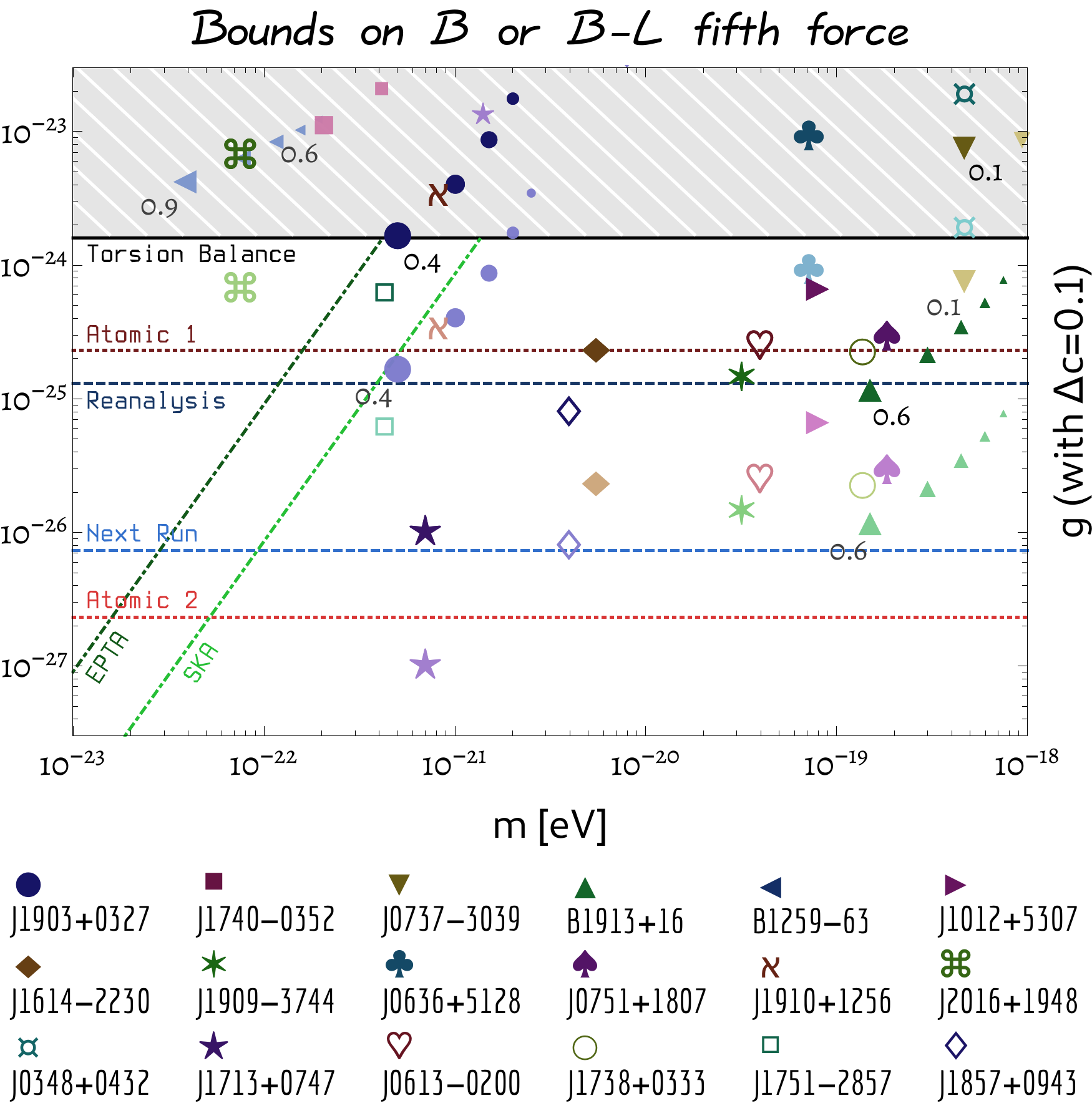}}
  \caption{Limits on the fifth force coupling \( g \) versus the vector fuzzy DM mass \( m \). Dark coloured symbols are the current bounds obtained from the corresponding systems with parameters given in Table~\ref{tab:frac}.  The same symbols in lighter colours show the constraints that would be obtained for the same systems were the precision on \(\dot{P}_b\) a factor of 10 higher.  The numerical labels give the eccentricity for eccentric, \(e\geq0.1\), systems.  The largest symbols refer to the first resonance \(N=1\), and the constraints for higher resonances (up to \(N=5\)) are shown with the same symbols but progressively smaller sizes.  The shaded region above the solid black line is excluded by torsion balance experiments ~\cite{Wagner:2012ui}.  For comparison, the plot includes the forecasted constraints obtained in ~\cite{Graham:2015ifn} from two different setups, using torsion pendulum (dashed dark blue line, and dashed light blue line), atom interferometry (dotted dark red line, and dotted light red line), and pulsar timing arrays (European Pulsar Timing Array, dot-dashed dark green line and Square Kilometer Array dot-dashed light green line).}\label{fig:direct}
\end{figure*}

In Fig.~\ref{fig:direct} we show the limits on the fifth force coupling \( g \) versus the vector fuzzy DM mass \( m \) that we obtain from the systems in Table~\ref{tab:frac}, assuming \( \Dc = 0.1 \), and ignoring an eventual suppression of the effect due to a coincidence in the values of the phase of the low-frequency modulation and the direction of the DM field\footnote{More precisely, we take the sum of the coefficient proportional to \(- s_\vth \cos(\gamma_l(t)+\vph)\) and \(- s_\vth \cos(\gamma_l(t)-\vph)\).  For near circular orbits, this is equivalent to assuming \( s_\vth \cos(\gamma_l(t)+\vph) =-1 \).}.  The limits are obtained by requiring that the magnitude of the effect of DM on \( \dot{P}_b\), be smaller than the error, \( \delta\dot{P}_b\), up to which it is known such effect is absent for those systems. For systems for which \( \dot{P}_b\) was measured, the value of \( \delta\dot{P}_b\) listed in Table~\ref{tab:frac} corresponds to the error on the intrinsic \( \dot{P}_b\) (that is, the measured value minus all the contributions of known effects), while for the others \( \delta\dot{P}_b\) represents an upper bound on \( \dot{P}_b\).

The dark, largest coloured symbols refer to the first harmonic \( N=1 \); we include the \( N=2 \) to \( N=5 \) higher harmonics which are displayed in progressively smaller copies of the same symbol\footnote{Note that for systems with small eccentricity the constraints from higher harmonics are too weak and, therefore, the corresponding symbols fall out of the range of the plot.}.  The symbols in lighter colours show what the constraints would be if the precision on the \(\dot{P}_b\) measurement improves by a factor of~10.  The numerical labels give the eccentricity for systems where \( e\geq0.1 \).  Included in the figure are also: (1) the constraints from torsion balance experiments~\cite{Wagner:2012ui} (solid black line, the excluded region is the shaded region above it); (2) the forecasted constraints from atom interferometry experiments with sensitivities of \( 10^{-13}\text{g}/\text{Hz}^{1/2} \) (dotted dark red line) and \( 10^{-15}\text{g}/\text{Hz}^{1/2} \) (dotted light red line)~\cite{Graham:2015ifn}; (3) the forecasted constraints from the reanalysis of torsion pendulum data (dashed dark blue line) and the next experimental run (dashed light blue line)~\cite{Graham:2015ifn}; (4) the sensitivities of the future European Pulsar Timing Array (EPTA, dot-dashed dark green line) and Square Kilometer Array (SKA, dot-dashed light green line)~\cite{Graham:2015ifn}.

 The main result of this section is that current pulsar timing data, for a wide range of masses, can already place the most stringent constraints on a \( B \) or \( (B-L) \) fifth force if this fifth force is carried by (fuzzy) DM.

As we already noticed, the typical measured values for the secular change in \( P_b \) can reach \( \dot{P}_b \lesssim 10^{-15} \text{s} \, \text{s}^{-1} \), and are expected to go down an order of magnitude with future data.  This means that fifth force couplings of the order \( g \sim 10^{-26} \) are within reach of current data, and will improve significantly in the near \( \cO{\text{year}} \) future.  Ideally, if \(\dot{P}_b \lesssim 10^{-16} \text{s} \, \text{s}^{-1}\) were achievable for long-period systems (\(P_b\gtrsim100 \text{d}\)), corresponding to \(m\lesssim10^{-21}\), we would be able to push the limit on \(g\) all the way down to \( g \sim 10^{-29} \) or even further.   Notice that  both the effect  in \Eq{eq:dir_dotp}  and the expected statistical uncertainty  of $\dot{P}_b$ scale with the orbital period as  $P_b^{4/3}$ \cite{Teukolsky1976,Damour:1991rd}.
Moreover, with the next generation of radio telescopes the number of binary systems suitable for timing analysis is expected to increase by a factor of~\(\sim10\), significantly covering the fuzzy DM mass range~\cite{Kramer:2015bea}.  Finally, we have the chance of detecting pulsar-black-hole binary systems (e.g.,~\cite{Kramer:2004hd,Keane:2014vja}), which would be ideal systems for testing the secular effect imprinted by the coupling between the DM field and the neutron star.

The same interaction Lagrangian \Eq{eq:lag-dir} can be used to describe the case of a dark photon; in this case either the photon is charged under a new \( U(1) \) which is carried by \( A_i \), or the \( A_i \) itself carries a very small electric charge.  This causes the two vector fields to mix and potentially generate a fifth force just as before.  However it is easy to see that this effect is unobservable since the binary members are practically electrically neutral.  This is so even ignoring any suppression due to plasma environment and assuming that \( q_k \) can be as large as \( \cO{1} \) for the masses we are interested in~\cite{Dubovsky:2015cca}.

A further possibility is to couple directly the vector field with the mass term via
\[
  M_k \rightarrow M_k(A) \deq M_k\left(1+\frac{A^2}{\Lambda^2}\right) \,,
\]
with \( \Lambda \) a mass scale that regulates the strength of the coupling.  In this case however vector fuzzy DM is in no way different from scalar fuzzy DM, which has been discussed in~\cite{Blas:2016ddr,Blas:2018} for the quadratic coupling between scalar field and mass.

%-------------------------------------------------------------------------------
\section{Discussion and conclusion}
\label{sec:end}
%-------------------------------------------------------------------------------

Even though in previous sections we have focussed on the secular variation of the orbital period, from the results we provide in the Appendix it is possible to work out the effects on all orbital parameters.  As emphasised in~\cite{Blas:2018}, it would be worth to perform such analysis to assess whether the constrains obtained only from \(\dot{P}_b\) can be improved.  In particular, as for the scalar field case, there are situations in which the secular variation of \(\dot{P}_b\) is negligible, but the secular drift for other parameters is not.

For instance, for the direct coupling of \Eq{eq:lag-dir}, from the knowledge of \(\dot{P}_b\), we cannot obtain constrains on masses in resonance with any \(N\geq2\) if the system has \(e\ll1\), see Fig.~\ref{fig:direct}.  However, as we show next, there is an effect on other orbital parameters that could be useful.  Indeed, using the definitions in \Eq{eq:qdef} and the properties in \Eqs{Jexpa1}{Jexpa2}, we have that in the limit \(e\to0\),
\begin{subequations}
  \begin{align}
    q_x(2e) \sim q_y(2e)		\sim &\, \frac{e}{2} \,,\\
    q_{xy}(2e) \sim q_{xx}(2e)	\sim &\, \frac{1}{2} \,.
  \end{align}
\end{subequations}
Then, the results collected in Panel~\ref{fig:eq2} for the secular variation of the orbital parameters reduce to
\begin{subequations}
  \begin{align}
    \avg{\dot e}		= &\, -\frac{q\sqrt{2\rhoDM}}{4a\omega_0 M_\odot} s_\vth c_{\gamma_l(t)+\vph} \deq \frac32 \frac{F^{SEP,eff}_y}{a\omega_0} \,, \label{eq:ex_dote}\\
     \avg{\dot\omega}	= &\, \frac{q\sqrt{2\rhoDM}}{4ae\omega_0 M_\odot} s_\vth s_{ \gamma_l(t)+\vph} \deq -\frac32 \frac{F^{SEP,eff}_x}{a\omega_0} \,, \label{eq:ex_dotom}
  \end{align}
\end{subequations}
with \(\avg{\dot{a}/{a}} = \avg{\dot\Om} = \avg{\dot\io} = \avg{\dot\ep_1} = 0\).  This effect is equivalent to the one obtained in~\cite{Damour:1991rq}, which arises when the strong equivalence principle (SEP) is violated\footnote{The effect discovered in~\cite{Damour:1991rq} is produced because the accelerations of the bodies in the gravitational field of the Galaxy are different from the acceleration \(\vec{g}\) of test bodies due to violations of the SEP.  It can be written as in \Eqs{eq:ex_dote}{eq:ex_dotom}, with \(\vec{F}^{SEP}=(\Delta_1-\Delta_2)\vec{g}\), where \(\Delta_k\) is the gravitational-to-inertial mass ratio.}, but with an effective \(\vec{F}^{SEP}\), given by\footnote{Notice that the magnitude of \(\vec{F}^{SEP,eff}\) is independent of time: \(|\vec{F}^{SEP,eff}|=q\sqrt{2\rhoDM} s_\vth/({6 M_\odot})\).}
\begin{align}
  \vec{F}^{SEP,eff} =	&\, -\frac{q\sqrt{2\rhoDM}}{6 M_\odot} s_\vth \left[s_{\gamma_l(t)+\vph}\hat{x} + c_{\gamma_l(t)+\vph}\hat{y}\right] \,. \label{SEPeff}
\end{align}

Analogously to what we have done with \(\dot{P}_b\), using now that the secular contribution to \({\dot e}\) should be smaller than the error \(\delta\dot e\), we obtain (assuming \(s_\vth=1\))
\begin{align} \label{avedot}
  g\frac{\Delta c}{0.1} \lesssim	&\, \delta\dot e \, \frac{40m_n a \omega_0} {\sqrt{2\rhoDM}} \sim 8.2\times 10^{-6} s\,\delta\dot e \,\left(\frac{M_T}{M_\odot}\right)^{1/3} \left(\frac{\text{d}}{P_b}\right)^{1/3} \,,
\end{align}
where in the numerical estimate we assumed \(\rhoDM = 0.3 \mbox{GeV}/\mbox{cm}^3\).  The  J1713+0747 system has \(\dot e=(-3\pm4)\times 10^{-18} s^{-1}\)~\cite{Zhu:2018etc}, which gives
\begin{align}
  g\frac{\Delta c}{0.1}\lesssim 8\times 10^{-24} \,, \,\, \mbox{for}~~~m\sim1.4\times 10^{-21} \mbox{eV} \,.
\end{align}
This upper bound is already competitive with the ones in Fig.~\ref{fig:direct} and, according to the analysis presented in~\cite{Freire:2012nb}, with an improvement in the TOA precision by a factor of~2, the bound is expected to be an order of magnitude smaller by 2030.  It is worth to recall here the advantages of using \(\dot e\) in comparison with \(\dot{P}_b\) for the purpose of constraining the fifth force~\cite{Damour:1991rd,Freire:2012nb}: while for many systems the error on \(\dot{P}_b\) is already dominated by the uncertainties in the value of the different contributions (mainly, the kinematic effect due to proper motion), the measured quantity \(\dot e\) is more robust since known contributions are expected to be subdominant with respect to the error of the timing measurement (which is expected to decrease with the time of observation, as \(T^{-3/2}\)).  Notice   that the amplitude of the effect on \( {\dot{P}_b}\) decreases with the orbital period as \(P_b^{4/3}\) while for \( {\dot e}\) it depletes only as \(P_b^{1/3}\) (see \Eq{eq:dir_dotp} and \Eq{avedot}, respectively). Notice also that while the scaling of the expected statistical uncertainty of \( {\dot{P}_b}\)   is the same  as the effect,  \(P_b^{4/3}\), the one of \( {\dot e}\) decreases faster, as \(P_b^{2/3}\) ~\cite{Teukolsky1976,Damour:1991rd}. Therefore, the bounds on \( {\dot e}\) could become more relevant for systems with shorter periods.    

It is interesting to see that, with this kind of systems, that is, when \(e\ll1\), the constraints obtained from \(\dot{P}_b\) refer to resonances with \(N=1\), while the ones derived here from \(\dot e\) correspond to \(N=2\).  This means that the different orbital parameters of the same system are probing, in a non-trivial way, different masses of the DM field.

We can compare the effective perturbation of \Eq{SEPeff} with the effect obtained in~\cite{Blas:2018} for a scalar DM field \(\Phi\).  The latter is present if the field is directly coupled to the bodies via the mass term, \(M_k \rightarrow M_k(\Phi) \deq M_k\left(1+\alpha_k {\Phi}\right)\), and the field gradient amounts to an effective DM velocity \( \vec{V} \) with respect to the barycenter of the binary.  The comparison is straightforward: after writing the contribution to  the scalar interaction that is linear in \(\vec{V}\), one can immediately see that the result is identical to that in \Eq{eq:lag-dir} if one defines an effective vector field \(\vec{A}^{\text{eff}}\deq-\Phi \vec{V}/|\vec{V}|\), and effective charges \(q_{k}^{\text{eff}} \deq{\alpha_k} |\vec{V}|{M_k}\).  Of course, the absence of any observation of a residual secular drift allows us to constrain both effects, but discriminating between them (which is considerably more challenging) would involve a combination of independent constrains on the different quantities.

Alongside scalars and vectors, fuzzy DM can potentially exist in the form of a spin-2 field, see~\cite{Marzola:2017lbt,Aoki:2017cnz}.  We plan to perform a careful evaluation of the distinctive features of the spin-2 fuzzy DM in pulsar timing observations in an upcoming work~\cite{LopezNacir:2018}.

%-------------------------------------------------------------------------------
\acknowledgments
%-------------------------------------------------------------------------------

FU wishes to thank A. Drago for useful correspondence.  FU is supported by the European Regional Development Fund (ESIF/ERDF) and the Czech Ministry of Education, Youth and Sports (MEYS) through Project CoGraDS - \verb|CZ.02.1.01/0.0/0.0/15_003/0000437|. DLN is supported by CONICET. DLN thanks D.~Blas, P.~Freire, and S.~Sibiryakov for useful discussions on related matters.

%-------------------------------------------------------------------------------
\appendix
\section{Appendix}
%-------------------------------------------------------------------------------

We collect here the useful formulas of Keplerian mechanics and the osculating orbits formalism.  More details can be found in~\cite{Danby:1970}.  Following the same notation as in~\cite{Blas:2018} we write down the Lagrange planetary equations,
\begin{subequations}
  \begin{align}
    \frac{\dot{a}}{a}	= &\, \frac{2}{\omega_0}\left\{\frac{e\sin\ta}{a\te} F_r + \frac{\te}{r} F_\ta \right\} \,, \label{eq:lag-a}\\
    \dot{e}				= &\, \frac{\te}{a\omega_0}\left\{(\cos\ta+\cos E) F_{\ta} + \sin\ta F_{r}\right\} \,, \label{eq:lag-e}\\
    \dot{\Om}			= &\, \frac{r\sin(\ta+\om)}{a^2\omega_0\te\sin\io} F_{z} \,, \label{eq:lag-Om}\\
    \dot{\io}			= &\, \frac{r\cos(\ta+\om)}{a^2\omega_0\te} F_{z} \,, \label{eq:lag-i}\\
    \dot{\vpi}			= &\, \frac{\te}{ae\omega_0}\left\{\left[1+\frac{r}{a\te^2}\right]\sin\ta F_{\ta} - \cos\ta F_{r} \right\} + 2\sin^2\left(\io/2\right)\dot{\Om} \,, \label{eq:lag-vpi}\\
    \dot{\ep_1}			= &\, -\frac{2r}{a^2\omega_0} F_{r} + \left(1-\te\right)\dot{\vpi} + 2\te\sin^2\left(\io/2\right)\dot{\Om} \,, \label{eq:lag-ep}
  \end{align}
\end{subequations}
in terms of the following six independent orbital elements: the semimajor axis \( a \) (not to be confused with the scale factor of the Universe), the orbital eccentricity \( e \), the longitude of the ascending node \( \Om \), the longitude of the periastron \( \vpi=\om+\Om \) (with \( \om \) the argument of the periastron, not to be confused with the orbital frequency \(\omega_0\)), the time of periastron \( t_0 \), and the inclination angle \( \io \) of the orbital plane with respect to the reference plane of the sky.  Here \( \ep_1=\omega_0(t-t_0)+\vpi-\int \dd t \,\omega_0 \), \( \omega_0=\sqrt{GM_T/a^3}=2\pi/P_b \), \( E \) is the eccentric anomaly that is defined by \( \omega_0(t-t_0)=E-e \sin E \).  We have also defined \( \te\deq\sqrt{1-e^2} \).  We use cartesian \( (x,y,z) \) and cylindric \( (r,\ta,z) \) coordinates in the orbital plane, and the overdot stands for a derivative with respect to time \( t \).  Therefore, \( \vec{r}\deq\hat{r}=r\cos{\ta}\hat{x}+r\sin\ta\hat{y} \), with \( \ta \) (not to be confused with the angle \( \vth \)) the angular position of \( M_1 \) with respect to the direction of the pericentre, \( \hat{x} \), and we have decomposed the perturbation as \( \vec{F}=F_r\hat{r}+F_{\ta}\hat{\ta}+F_z\hat{z} \).  The expressions of the components of \( \vec{F} \) or a generic vector in the \( (X,Y,Z) \) coordinates can be found in~\cite{Poisson:2014}.

The orbit can be expanded in a series as:
\begin{align}\label{eq:kepexp}
  \begin{cases}
    x/a		&  = (x0/a) + \sum q_x(ne) \cos(\nnut) \\
    y/a		&  = (y0/a) + \sum q_y(ne) \sin(\nnut) \\
    r/a		&   = (r0/a) - \sum q_r(ne) \cos(\nnut) \\
    (x/a)^2	&  = (x0/a)^2 + \sum q_{xx}(ne) \cos(\nnut) \\
    (y/a)^2	&  = (y0/a)^2 + \sum q_{yy}(ne) \cos(\nnut) \\
    xy/a^2	&  = (x0y0/a)^2 + \sum q_{xy}(ne) \sin(\nnut) \\
    (r/a)^2	&  = (r0/a)^2 - \sum q_{rr}(ne) \cos(\nnut)
  \end{cases} \,,
\end{align}
where the sums run over \( n\in[1,\infty)\) and the zeroth terms are not necessary as in the end only the resonant harmonics will be relevant.  The expansion coefficients are 
\begin{align}\label{eq:qdef}
  \begin{cases}
    q_x(ne)			& 	\deq 2J'_n(ne)/n \\
    q_y(ne)			& 	\deq 2\sqrt{1-e^2}/e J_n(ne)/n \\ 
    q_r(ne)			& 	\deq 2e J'_n(ne)/n \\ 
    n q_{xx}(ne)	& 	\deq J_{n-2}(ne) - J_{n+2}(ne) - 2e\left[ J_{n-1}(ne) - J_{n+1}(ne) \right] \\
					& 	= 4J_{n}' (ne)\frac{(1-e^2)}{e}-\frac{4 J_n(ne)}{n e^2} \\
    n q_{yy}(ne)	& 	\deq (1-e^2)\left[ J_{n+2}(ne) - J_{n-2}(ne) \right] \\
					& 	= - n q_{xx}(ne) - 4J_n(ne)/n \\
    n q_{xy}(ne)	& 	\deq \sqrt{1-e^2}\left[ - 2J_n(ne) + J_{n+2}(ne) + J_{n-2}(ne) \right] \\
					& 	= 4\sqrt{1-e^2}\left[ \, J_{n}(ne)\frac{(1-e^2)}{e^2} - \frac{J_n'(ne)}{n e} \right] \\
    n q_{rr}(ne)	& 	\deq 4J_n(ne)/n^2 \\
					& 	= -n^2\left(q_{xx}(ne)+q_{yy}(ne)\right)
  \end{cases}\,,
\end{align}
where the \( J_n(z) \) are Bessel functions of the first kind.  Some useful relations can be found among expansion coefficients:
\begin{subequations}
  \begin{align}
    q_x(ne)	= &\, -\frac{1}{2e\te^2} \left[ \te^2 q_{xx}(ne) - q_{yy}(ne) \right] \,, \nn\\
	q_y(ne) = &\, -\frac{\te n}{2e} \left[ q_{xx}(ne) + q_{yy}(ne) \right] \,. \nn
  \end{align}
\end{subequations}
With the use of the expansion of the Bessel function and its derivative for small values of \(e\), 
\begin{subequations}
  \begin{align}
    J_N(Ne)		& = \frac{\left(Ne/2\right)^N}{\Gamma[N+1]}\left[1+\cO{e^2}\right] \,, \label{Jexpa1}\\
    J'_N(Ne)	& = \frac{J_N(N e)}{e}\left[1+\cO{e^2}\right] \,, \label{Jexpa2}
  \end{align}
\end{subequations}
where \(\Gamma[x]\) is the Gamma function, we obtain that for nearly circular orbits  
\begin{subequations}
  \begin{align}
    q_y(ne)	& = \frac{\left(Ne/2\right)^{N-1}}{\Gamma[N+1]}\left[1+\cO{e^2}\right] \,,\label{qJexpa1}\\
    q_x(ne)	& = q_y(ne) \left[1+\cO{e^2}\right] \,.\label{qJexpa2}
  \end{align}
\end{subequations}

In the Panels~\ref{fig:eq1} and~\ref{fig:eq2} we collect the secular changes of all orbital parameters for the cases of the gravitational force \Eq{eq:pertb} and the direct coupling case \Eq{eq:force-dir}, respectively.

\begin{panel*}[htbp]
\caption{Secular changes of all six orbital parameters for the gravitational force \Eq{eq:pertb}.  All the \( q_{ij} \) are functions of \( Ne \), and the frequency gap is defined as \( \denu \deq 2m - N\omega_0 \), and we are using the notation defined in \Eq{shortcut}.}
\begin{subequations}
  \begin{align}
    \avg{\frac{\dot{a}}{a}}	= &\, \frac{2\pi G\rhoDM N}{\omega_0} \left\{ \left[\left(4 s_\vth^2 c_\vph^2 - 1 \right)q_{xx}+\left(4 s_\vth^2 s_\vph^2 - 1 \right)q_{yy}\right] s_{\gamma_g(t)} + 4 s_\vth^2 s_{2\vph} q_{xy} c_{\gamma_g(t)} \right\} \label{eq:lag-a2}\\
    \avg{\dot e}			= &\, \frac{\pi G\rhoDM\te}{e\omega_0} \left\{\left[\left(4 s_\vth^2 c_\vph^2 - 1 \right)\te Nq_{xx} + \left(4 s_\vth^2 s_\vph^2 - 1 \right)\te Nq_{yy} - 8 s_\vth^2 c_{2\vph} q_{xy} \right] s_{\gamma_g(t)} \right. \nn\\
							  &\, \left. - 4 s_\vth^2 s_{2\vph} \left[ q_{xx} - q_{yy} - \te Nq_{xy} \right] c_{\gamma_g(t)} \right\} \label{eq:lag-e2}\\
    \avg{\dot\Om}			= &\, \frac{4\pi G\rhoDM \csc\!\io}{\te\omega_0} s_{2\vth} \left\{ -c_{\vph-\om} q_{xy} s_{\gamma_g(t)} + \left[s_\om c_\vph q_{xx}+c_\om s_\vph q_{yy}\right] c_{\gamma_g(t)} \right\} \label{eq:lag-Om2}\\
    \avg{\dot\io}			= &\, \frac{4\pi G\rhoDM}{\te\omega_0} s_{2\vth} \left\{ s_{\om-\vph} q_{xy} s_{\gamma_g(t)} + \left[c_\om c_\vph q_{xx}-s_\om s_\vph q_{yy}\right] c_{\gamma_g(t)} \right\} \label{eq:lag-i2}\\
    \avg{\dot\vpi}			= &\, 2\sin^2\left(\io/2\right)\avg{\dot\Om} + \frac{\pi G\rhoDM}{e^2\te\omega_0} \left\{ 4s_\vth^2 s_{2\vph} \left[ (1+\te^2)q_{xy} - \te^3 N\left(2q_{xx}+q_{yy}\right) \right] s_{\gamma_g(t)} \right.  \nn\\
							  &\, \left. +\left[\left(4s_\vth^2 c_\vph^2 - 1 \right)\te^2 q_{xx} + \left(4s_\vth^2 (c_\vph^2+2c_{2\vph}) - 1 \right) q_{yy} - 4s_{\vth}^2 c_{2\vph} \te^3 N q_{xy} \right] c_{\gamma_g(t)} \right\} \label{eq:lag-vpi2}\\
    \avg{\dot\ep_1}			= &\, \frac{4\pi G\rhoDM}{\omega_0} \left\{2 s_\vth^2 s_{2\vph} q_{xy} s_{\gamma_g(t)} - \left[\left(4 s_\vth^2 c_\vph^2 - 1 \right)q_{xx}+\left(4 s_\vth^2 s_\vph^2 - 1 \right)q_{yy}\right] c_{\gamma_g(t)} \right\} \nn\\
							  &\, + \left(1-\te\right)\avg{\dot\vpi} + 2\te\sin^2\left(\io/2\right)\avg{\dot\Om} \label{eq:lag-ep2}
  \end{align}
\end{subequations}\\
\centering\rule[12pt]{0.485\textwidth}{0.1pt}
\label{fig:eq1}
\end{panel*}

\begin{panel*}[htbp]
\caption{Secular changes of all six orbital parameters for the direct coupling case, where the perturbation is given by \Eq{eq:force-dir}.  All the \( q_{ij} \) are functions of \( Ne \), and the frequency gap is defined as \( \denu' \deq m-N\omega_0 \) (notice the factor of 2 difference from \( \denu \)), and we are using the notation defined in \Eq{shortcut}.}
\begin{subequations}
  \begin{align}
    \avg{\frac{\dot{a}}{a}}	= &\, \frac{q\sqrt{2\rhoDM}N}{a\omega_0 M_\odot} s_\vth \left\{ q_y s_\vph s_{\gamma_l(t)} - q_x c_\vph c_{\gamma_l(t)} \right\} \\
    \avg{\dot e}				= &\, -\frac{q\sqrt{2\rhoDM}\te}{2ae\omega_0 M_\odot} s_\vth \left\{\left[q_x-\te N q_y\right] s_\vph s_{\gamma_l(t)} - \left[q_y-\te N q_x\right] c_\vph c_{\gamma_l(t)} \right\} \\
    \avg{\dot\Om}				= &\, \frac{q\sqrt{2\rhoDM}\csc\!\io}{2a\te\omega_0 M_\odot} c_\vth \left\{ s_\om q_x s_{\gamma_l(t)} + c_\om q_y c_{\gamma_l(t)} \right\} \\
    \avg{\dot\io}				= &\, \frac{q\sqrt{2\rhoDM}}{2a\te\omega_0 M_\odot} c_\vth \left\{ c_\om q_x s_{\gamma_l(t)} - s_\om q_y c_{\gamma_l(t)} \right\} \\
    \avg{\dot\vpi}			= &\, \frac{q\sqrt{2\rhoDM}N}{4ae\omega_0 M_\odot} s_\vth \left\{ c_\vph q_{xy} s_{\gamma_l(t)} + s_\vph q_{xx} c_{\gamma_l(t)} \right\} + 2\sin^2\left(\io/2\right)\avg{\dot{\Om}} \\
    \avg{\dot\ep_1}			= &\, -\frac{q\sqrt{2\rhoDM}}{a\omega_0 M_\odot} s_\vth \left\{ c_\vph q_x s_{\gamma_l(t)} + s_\vph q_y c_{\gamma_l(t)} \right\} + \left(1-\te\right)\avg{\dot{\vpi}} + 2\te\sin^2\left(\io/2\right)\avg{\dot{\Om}}
  \end{align}
\end{subequations}\\
\centering\rule[12pt]{0.485\textwidth}{0.1pt}
\label{fig:eq2}
\end{panel*}

Table~\ref{tab:frac} lists all the binary systems that we have used in this study, alongside their relevant properties.
\begin{table*}[htbp]
  \centering
  \begin{tabular}[h]{lccccccr}
    \toprule
    Name & \(M_1\) [\(M_\odot\)] & \(M_2\) [\(M_\odot\)] & \(e\) & \(P_b\) [d] & \(\dot{P}_b~[\text{s}\,\text{s}^{-1}]\) & \(\delta\dot{P}_b~[\text{s}\,\text{s}^{-1}]\) & References \tabularnewline
    \midrule
    J1903+0327	& 1.03		& 1.67		& 0.44		& 95		& -6.4e-11		& 3.1e-11	& \cite{Freire:2010tf} \tabularnewline
	J1740-3052	& 20		& 1.4		& 0.58		& 231		& 3e-9			& 3e-9		& \cite{Madsen:2012rs} \tabularnewline
	J0737-3039	& 1.249		& 1.338		& 0.088		& 0.1022	& -1.252e-12	& 0.017e-12	& \cite{Wex:2014nva} \tabularnewline
	B1913+16	& 1.39		& 1.44		& 0.62		& 0.32		& -2.423e-12	& 0.001e-12	& \cite{Wex:2014nva} \tabularnewline
	B1259-63	& 24		& 30		& 0.87		& 1237		& 1.4e-8		& 0.7e-8	& \cite{Shannon:2013dpa} \tabularnewline
	J1012+5307	& 0.10\dg	& 1.2\ddg	& 1.3e-6	& 0.60		& 8.1e-14		& 2.0e-14	& \cite{Arzoumanian:2017puf,Desvignes:2016yex} \tabularnewline
	J1614-2230	& 0.49		& 1.9		& NA		& 8.7		& 1.7e-12		& 0.2e-12	& \cite{Arzoumanian:2017puf} \tabularnewline
	J1909-3744	& 0.21		& 1.5		& 1.2e-7	& 1.5		& 5.02e-13		& 0.13e-13	& \cite{Arzoumanian:2017puf,Desvignes:2016yex} \tabularnewline
	J0636+5128	& 0.007\dg	& 1.4\ddg	& 2.2e-5	& 0.67		& 2.5e-12		& 0.3e-12	& \cite{Arzoumanian:2017puf,Stovall:2014gua} \tabularnewline
	J0751+1807	& 0.16		& 1.6		& 3.3e-6	& 0.26		& -3.50e-14		& 0.25e-14	& \cite{Desvignes:2016yex} \tabularnewline
	J1910+1256	& 0.3		& 1.6		& 2.3e-4	& 58		& -2e-11		& 4e-11		& \cite{Gonzalez:2011kt} \tabularnewline
	J2016+1948	& 0.45		& 1.0		& 1.5e-3	& 635		& -1e-9			& 2e-9		& \cite{Gonzalez:2011kt} \tabularnewline
	J0348+0432	& 0.17		& 2.0		& 2.4e-6	& 0.1024	& -2.73e-13		& 0.45e-13	& \cite{Antoniadis:2013pzd} \tabularnewline
	J1713+0747	& 0.29		& 1.33		& 7.5e-5	& 68		& 3.4e-13		& 1.5e-13	& \cite{Zhu:2018etc} \tabularnewline
	J0613-0200	& 0.12\dg	& 1.2\ddg	& 5.4e-6	& 1.2		& 5.4e-14		& 1.8e-14	& \cite{Arzoumanian:2017puf,Desvignes:2016yex} \tabularnewline
	J1738+0333	& 0.18		& 1.46		& 3.4e-7	& 0.35		& -1.7e-14		& 0.3e-14	& \cite{Freire:2012mg} \tabularnewline
	J1751-2857	& 0.18\dg	& 1.2\ddg	& 1.3e-4	& 111		& 1.8e-11		& 1.8e-11	& \cite{Desvignes:2016yex,Caputo:2017zqh} \tabularnewline
	J1857+0943	& 0.27\dg	& 1.2\ddg	& 2.2e-4	& 12		& 1.2e-13		& 1.2e-13	& \cite{Desvignes:2016yex,Caputo:2017zqh} \tabularnewline
    \bottomrule
  \end{tabular}
  \caption{List of binary systems used in this study.  The columns are: (1) the name of the binary; (2) the mass of the companion in \(M_\odot\) units (if only the minimum value is available we denote this with a \dag); (3) the mass of the pulsar in \(M_\odot\) units (assumed values are indicated with a \ddag); (4) the orbital eccentricity; (5) the binary period in days; (6) the period derivative in \(\text{s}\,\text{s}^{-1}\); (7) the upper limit or error on the period derivative, also in \(\text{s}\,\text{s}^{-1}\); (8) the references.  In Fig.~\ref{fig:direct} we have assumed zero eccentricity \(e=0\) for J1614-2230.}\label{tab:frac}
\end{table*}

\clearpage
\bibliographystyle{hieeetr}
\bibliography{biblio}

\end{document}